\documentclass[cameraready]{Interspeech}

\usepackage{amsmath,graphicx,hyperref, multirow}

\usepackage{amsmath,amssymb,amsfonts}
\usepackage{algorithmic}
\usepackage{graphicx}
\usepackage{textcomp}
\usepackage{xcolor}
\usepackage{caption}
\usepackage{booktabs}
\usepackage{arydshln}
\usepackage{makecell}
\title{Toward Open-Set Speaker Attribute Prediction with Keyword-Appended LLM Embeddings}

\author[affiliation={1}]{Byoungjun}{So}
\author[affiliation={1}, orcid=0009-0008-4392-0583]{Jaejun}{Lee}
\author[affiliation={1,2,3}, orcid=0000-0002-4210-0312]{Kyogu}{Lee}

\email{nirv1120@snu.ac.kr, jjlee0721@snu.ac.kr, kglee@snu.ac.kr}
\address{
    $^1$ Department of Intelligence and Information, Seoul National University\\
    $^2$ Interdisciplinary Program in Artificial Intelligence, Seoul National University\\
    $^3$ Artificial Intelligence Institute, Seoul National University, Republic of Korea
}

\email{\{nirv1120, jjlee0721, kglee\}@snu.ac.kr}

\keywords{Speaker attribute prediction, LLM}

\usepackage{comment}

\begin{document}

\maketitle

\begin{abstract}
    Understanding speaker attributes is crucial for voice-related applications, yet conventional approaches rely on fixed categorical labels, lacking semantic richness and zero-shot generalizability. We propose a novel framework for open-set speaker attribute prediction leveraging Large Language Model (LLM) embeddings to represent attributes in a continuous semantic space. To bridge the cross-modal gap, we introduce a keyword-appending strategy that structures broad semantic representations into a compact, discriminative manifold. Furthermore, we employ a top-$k$ negative loss to establish robust decision boundaries in crowded semantic regions. Experimental results on LibriTTS-P demonstrate that our method outperforms closed-set benchmarks and generalizes effectively to unseen synonyms. 
    Geometric analysis suggests that our strategies regularize the embedding manifold, balancing semantic cohesion with predictive clarity.
\end{abstract}

\section{Introduction}
Modeling speaker identity is an essential component of modern speech technologies, enabling applications ranging from speaker recognition to multi-speaker text-to-speech (TTS) and voice conversion (VC).

Most existing approaches extract speaker information using frameworks that rely on intermediate representations from pretrained speaker verification networks~\cite{desplanques2020ecapa, chen2022wavlm, resemblyzer}. However, these representations are essentially black-box embeddings, offering little interpretability regarding specific explainable speaker attributes (\textit{e.g.,} \textit{bright}, \textit{masculine}, or \textit{sharp}). 

Speaker attribute prediction is a newly emerging approach that explicitly maps acoustic speaker input to explainable voice-related linguistic keywords. While this approach provides a more transparent and interpretable speaker representation, previous research~\cite{lee25e_interspeech} relies on predefined attribute labels implemented within a closed-set multi-label classification framework. Such a rigid design lacks generalizability to attributes outside the dataset and fails to capture the rich semantic relationships and subtle gradations between them, thereby limiting the model's ability to extend to unseen attributes.

To overcome these limitations, we propose a novel open-set speaker attribute prediction framework. Our approach leverages Large Language Model (LLM) embeddings to map speaker attributes into a continuous semantic space, moving beyond fixed categorical labels. To effectively bridge the modality gap between acoustic speaker characteristics and the voice-semantic textual space, we introduce a latent space alignment strategy with keyword-appending and top-$k$ negative penalization, which sharpens the embedding space by shifting raw labels into a more compact and structured manifold.
Extensive experiments reveal that our approach not only enhances performance in closed-set attribute prediction but also demonstrates robust capability in identifying open-set attributes, such as synonyms. Furthermore, through rigorous geometric analysis, we show that attribute prediction performance is strongly correlated with the manifold structure spanned by the LLM-based embeddings. This provides empirical evidence that aligning the semantic latent space is crucial for achieving robust and generalizable speaker attribute prediction.

\section{Related Works}
\subsection{Speaker Attribute Prediction}
Conventional approaches to leveraging speaker information rely on intermediate embedding representations from speaker verification networks such as ECAPA-TDNN~\cite{desplanques2020ecapa}, WavLM-TDNN~\cite{chen2022wavlm}, and Resemblyzer~\cite{resemblyzer}. These speaker embeddings are widely used in recent speaker recognition tasks~\cite{jakubec2024deep, sharma2024milestones} or as conditional information for multi-speaker speech synthesis frameworks~\cite{zhangvevo, wang2025discl}. Furthermore, they are extensively employed to assess speaker information imprinting in synthesized speech~\cite{gusev2024improvement, lee2024hear, guo2024xe}. Although these approaches are deeply integrated into speech-related research, they lack interpretability due to their reliance on black-box representations.

Recent research has shifted toward explicit speaker attributes. The LibriTTS-P dataset~\cite{kawamura2024libritts} facilitated this by providing interpretable attributes paired with specific intensity annotations (\textit{e.g.,} \textit{slightly bright}, \textit{very halting}, or \textit{sharp}). This dataset has been widely adopted in conditional TTS studies, where the goal is to model user-friendly speaker identity conditions~\cite{guo2023prompttts, zhang2023promptspeaker, lengprompttts}. Building on the LibriTTS-P dataset, the speaker attribute prediction task has emerged~\cite{lee25e_interspeech}, formulating the problem as a closed-set multi-label classification to map speech into a predefined categorical attribute distribution. While this has helped bridge the gap between black-box representations and explainable attributes, it remains restricted to a predefined attribute palette, limiting its generalizability. This limitation motivates us to extend the task into an open-set speaker attribute prediction framework.

\subsection{LLM-Based Open-Set Prediction}
Building upon the rapid advancements in Large Language Models (LLMs)~\cite{achiam2023gpt, team2023gemini, grattafiori2024llama}, several studies have extended conventional closed-set classification into continuous semantic spaces. Notable examples include zero-shot generalization in music emotion recognition~\cite{liu2024leveraging} and image classification via the DeViSE framework~\cite{frome2013devise}. However, mapping attributes directly into a pre-trained LLM embedding space often results in complex regions, leading to semantic ambiguity between closely related descriptors.

To address this, margin-based metric learning and hard negative mining—such as Triplet Loss~\cite{schroff2015facenet} and InfoNCE~\cite{oord2018representation}—are widely employed to push apart confusing samples and refine the embedding manifold. Drawing on these principles, we propose a top-$k$ negative penalization method integrated with a keyword-appending strategy. This approach effectively reshapes the latent space to form speaker-attribute-specific LLM embeddings, ensuring a more discriminative and structured manifold for robust attribute prediction.

\section{Methods} \label{sec:method}
In this section, we describe the proposed open-set speaker attribute prediction framework. We first introduce our approach to leveraging LLM-based attribute embeddings ($\mathbf{e}$) and a keyword-appending strategy to construct a compact embedding space. Subsequently, we detail the top-$k$ negative penalization method, which structures the embedding space to be more semantically discriminative.
\subsection{LLM-Based Speaker Attribute Prediction}
Unlike previous work~\cite{lee25e_interspeech}, which employed binary cross-entropy (BCE) loss for closed-set multi-label classification, we utilize LLM embeddings as continuous targets. We adopt a cosine similarity-based objective to align predicted acoustic representations with LLM-derived attribute embeddings.
To account for the intensity of each attribute in speech, we introduce a weighted cosine similarity loss ($\mathcal{L}_{\text{wcos}}$), where the weight $w_j$ corresponds to the annotated intensity of the $j$-th attribute for a given audio sample. The training objective of our prediction model $f$ is defined as:
\begin{equation}
\mathcal{L}_{\text{wcos}} = \frac{\sum_{j \in \mathcal{P}} w_{j} \cdot (1 - \cos(f(x), \mathbf{e}_j))}{\sum_{j \in \mathcal{P}} w_{j}},
\label{eq:wcos}
\end{equation}
where $f(x)$ denotes the output representation of the prediction model for input audio $x$, $\mathbf{e}_{j}$ is the target LLM embedding of the $j$-th attribute, $\cos(\mathbf{a}, \mathbf{b}) = \frac{\mathbf{a}^{\top} \mathbf{b}}{|\mathbf{a}| |\mathbf{b}|}$ is the cosine similarity, and $\mathcal{P}$ denotes the set of positive attributes annotated for the sample.
By weighting the similarity according to the annotated intensity of each attribute, we encourage the predicted representation to be located in the appropriate region of the embedding space, such that its distance to each attribute embedding reflects the degree to which the corresponding voice characteristic is expressed in the input speech.

\subsection{Keyword Appending Strategy}
The speaker attributes obtained from LibriTTS-P~\cite{kawamura2024libritts}, as listed in Table~\ref{tab:attribute_synonyms}, are carefully designed to represent vocal characteristics. However, several attributes exhibit semantic ambiguity when considered outside the context of voice. For instance, the term \textit{`cute'} is not exclusively associated with speech; it frequently refers to other domains such as facial appearance or behavior. This ambiguity is even more pronounced for terms like \textit{`sweet'}, which spans a wide variety of contexts. In contrast, attributes such as \textit{`nasal'} or \textit{`muffled'} are more inherently tied to vocal quality. This discrepancy can lead to a sporadically structured embedding space, making it difficult for the prediction model to align acoustic features with their corresponding semantic representations.

To mitigate this, we introduce a keyword-appending strategy—a simple yet effective method to ground the attributes in the common domain. We append a domain-specific keyword (\textit{e.g.,} \textit{`speech'}, \textit{`voice'}) to each attribute (e.g., \textit{`cute'} becomes \textit{`cute speech'}, \textit{`nasal'} becomes \textit{`nasal speech'}). This ensures that the resulting LLM embeddings are more domain-specific, leading to a more compact and structured embedding manifold. By narrowing the semantic scope, this strategy facilitates easier cross-modal alignment between acoustic features and textual representations.

\begin{table}[t!]
\caption{The ground-truth attributes from the LibriTTS-P~\cite{kawamura2024libritts} dataset and their corresponding synonyms generated via Gemini 3.1 Pro~\cite{deepmind2026gemini31} for validating zero-shot attribute prediction.}
\centering
\resizebox{0.4\textwidth}{!}{
\begin{tabular}{|l l|l l|}
\hline
\textbf{Attribute} & \textbf{Synonym} & \textbf{Attribute} & \textbf{Synonym} \\
\hline
adult-like & grown-up & modest & humble \\
bright & brilliant & muffled & muted \\
calm & serene & nasal & twangy \\
clear & lucid & old & aged \\
cool & chill & powerful & forceful \\
cute & endearing & raspy & hoarse \\
dark & gloomy & reassuring & comforting \\
elegant & graceful & refreshing & revitalizing \\
feminine & womanly & relaxed & unhurried \\
fluent & eloquent & sexy & seductive \\
friendly & amiable & sharp & keen \\
gender-neutral & androgynous & sincere & genuine \\
halting & hesitant & soft & mild \\
hard & tough & strict & rigorous \\
intellectual & academic & sweet & sugary \\
intense & profound & tensed & tight \\
kind & gentle & thick & heavy \\
light & airy & thin & slim \\
lively & vivid & unique & distinctive \\
masculine & manly & weak & fragile \\
mature & ripe & wild & savage \\
middle-aged & elderly & young & youthful \\
\hline
\end{tabular}
}
\vspace{-4mm}
\label{tab:attribute_synonyms}
\end{table}

\subsection{Structuring Manifold via Top-$k$ Negative Penalization}
While the weighted cosine similarity loss ($\mathcal{L}_{\text{wcos}}$) aligns acoustic features with continuous LLM embeddings, and the keyword-appending strategy facilitates a domain-specific compact manifold, such semantic alignment can lead to a subspace crowding phenomenon. In these dense regions, semantically proximal attributes become closely packed, causing the model to easily confuse adjacent labels and degrading its discriminative power. To counteract this crowding and secure clear decision margins against irrelevant attributes, we adapt a top-$k$ negative loss ($\mathcal{L}_{\text{negk}}$).
For a given sample, we first compute a dynamic margin anchor $a$, defined as the weighted average cosine similarity over all positive attributes:
\begin{equation}
a = \frac{\sum_{j \in \mathcal{P}} w_{j} \cdot \cos(f(x), \mathbf{e}_j)}{\sum_{j \in \mathcal{P}} w_{j}}.
\label{eq:anchor}
\end{equation}
Using this anchor, we identify the set of top-$k$ negative attributes $\mathcal{N}$ that exhibit the highest cosine similarity to the prediction $f(x)$. We then penalize the model if the similarity of any attribute in $\mathcal{N}$ exceeds the margin-adjusted anchor $a-m$. Specifically, we employ the softplus function to softly penalize margin violations, ensuring training stability by providing smooth, non-zero gradients even near the margin threshold. This encourages the model to maintain a meaningful distance between positive attributes and semantically confusing negatives. The top-$k$ negative loss $\mathcal{L}_{\text{negk}}$ is formulated as:
\begin{equation}
\mathcal{L}_{\text{negk}} = \frac{1}{k} \sum_{j \in \mathcal{N}} \text{softplus}(\cos(f(x), \mathbf{e}_j) - (a - m)).
\label{eq:negk}
\end{equation}
where $\text{softplus}(x) = \ln(1 + \exp(x))$.

\subsection{Final Optimization Formulation}
The overall training objective for our speaker attribute prediction model is formulated as follows:
\begin{equation}
    \mathcal{L}_{\text{total}} = \mathcal{L}_{\text{wcos}} + \lambda_{\text{negk}} \mathcal{L}_{\text{negk}} 
\end{equation}
where $\lambda_{\text{negk}}$ is a hyperparameter that controls the relative contribution of the top\mbox{-}$k$ negative loss. By integrating $\mathcal{L}_{\text{negk}}$ with $\mathcal{L}_{\text{wcos}}$, our framework effectively refines the embedding space—specifically the keyword-appended compact space—into a more discriminative and structured manifold.
This approach not only extends speaker attribute prediction to an open-set setting but also ensures a more robust mapping from acoustic features to the textual space. By constructing a semantically compact manifold with clear marginal separation, our framework achieves the stability and discriminative power necessary for effective cross-modal alignment.

\begin{table}[t]
\centering
\caption{Micro-averaged $\mathrm{\textit{F}}_1$ scores for closed-set speaker attribute prediction.}
\label{tab:vo-ve-vs-proposed}
\resizebox{0.8\columnwidth}{!}{%
\begin{tabular}{ccc}
\toprule
\textbf{Threshold $\tau$} & \textbf{Benchmark~\cite{lee25e_interspeech}} & \textbf{Proposed model} \\
\midrule
0.2 & 0.6645 $\pm$ 0.0686 & 0.7625 $\pm$ 0.0482 \\
0.4 & 0.6908 $\pm$ 0.0719 & 0.7625 $\pm$ 0.0482 \\
0.6 & 0.6415 $\pm$ 0.0662 & 0.7625 $\pm$ 0.0482 \\
0.8 & 0.7286 $\pm$ 0.0489 & 0.7380 $\pm$ 0.0407 \\
\bottomrule
\end{tabular}
}
\vspace{-2mm}
\end{table}

\section{Experiments}
\subsection{Datasets}
For training and evaluation, we utilized the LibriTTS-P dataset~\cite{kawamura2024libritts}, which is currently the only open-source corpus providing speaker-wise attribute labels.
This dataset is built upon 
% LibriTTS-R~\cite{koizumi2023libritts}, a restored high-quality version of
the widely used LibriTTS corpus~\cite{zen2019libritts}.
The corpus includes speech from 2,443 speakers, where three annotators provided multi-label annotations across 44 distinct voice attribute categories (as shown in Table~\ref{tab:attribute_synonyms}). Each attribute is further annotated with varying intensity levels: \textit{very}, \textit{normal}, and \textit{slightly}. We adhered to the predefined train, validation, and test splits.

\subsection{Implementation Details}
To ensure a fair evaluation of our proposed framework, we adopt the ECAPA-TDNN architecture~\cite{desplanques2020ecapa} as the model backbone, following the benchmark established in~\cite{lee25e_interspeech}. The output dimension of the final prediction layer is set to 2,880 to align with the embedding size of GPT-OSS-20B~\cite{openai2025gptoss}, the recently released open-weights model from OpenAI.

For the three intensity levels of the voice attributes—\textit{very}, \textit{normal}, and \textit{slightly}—we assigned numerical weights ($w$) of 1.5, 1.0, and 0.5, respectively. Unannotated attributes were assigned a weight of 0. To derive the final target weight for each speaker, we calculated the mean of the weights provided by the three annotators and applied normalization. It should be noted that all speech samples belonging to the same speaker share the same attribute weights.
The hyperparameters for the loss functions were set as follows: the margin $m\!=\!0.2$, the scaling factor $\lambda_{\text{negk}}\!=\!0.5$, and the number of negatives $k\!=\!1$.

\subsection{Evaluation Settings}
We compared our proposed framework against the primary benchmark~\cite{lee25e_interspeech} using its official implementation\footnote{https://github.com/jaejunL/vove}. This benchmark is a closed-set multi-label classification model trained to predict predefined discrete indices for 44 attribute classes.

To evaluate the performance of speaker attribute prediction, we report micro-averaged $F_1$ scores. Given the multi-label nature of the task, predictions are determined by applying decision thresholds to the cosine similarity scores. We evaluate performance across four thresholds: 0.2, 0.4, 0.6, and 0.8. For the benchmark, which utilizes binary cross-entropy (BCE) loss, the $F_1$ scores are calculated based on the sigmoid activation outputs relative to these thresholds.

\begin{table}[t!]
\centering
\caption{Micro-averaged $\mathrm{\textit{F}}_1$ scores with various keywords in zero-shot synonym prediction at various thresholds $\tau$.}
\label{tab:transposed-f1}
\resizebox{0.48\textwidth}{!}{
\begin{tabular}{ccccccc}
\toprule
\multirow{2}{*}{\textbf{$\tau$}} & \multicolumn{5}{c}{\textbf{Proposed (with keyword appending)}} & \multirow{2}{*}{\makecell[c]{\textbf{No} \\ \textbf{keyword}}} \\ \cmidrule(lr){2-6}
 & \textbf{--speech} & \textbf{--voice} & \textbf{--face} & \textbf{--man} & \textbf{--apple} & \\ \midrule
0.2 & 0.7629 & 0.7621 & 0.7615 & 0.7584 & 0.7614 & 0.7628 \\
0.4 & 0.7627 & 0.7601 & 0.7613 & 0.7581 & 0.7613 & 0.7628 \\
0.6 & 0.7582 & 0.7520 & 0.7591 & 0.7569 & 0.7612 & 0.7628 \\
0.8 & 0.6920 & 0.6763 & 0.7193 & 0.7126 & 0.7606 & 0.0018 \\ \bottomrule
\end{tabular}
}
\vspace{-2mm}
\end{table}

\section{Results}
In this section, we present a comprehensive evaluation of our proposed framework. First, we compare our model against the benchmark in a closed-set speaker attribute prediction task. Second, we demonstrate the open-set capability of our model through a zero-shot synonym attribute prediction task, highlighting its ability to generalize beyond predefined labels. Finally, we provide an in-depth geometric analysis to verify how the keyword-appending strategy and top-$k$ negative penalization effectively structure the embedding space.

\subsection{Closed-Set Speaker Attribute Prediction}
As shown in Table~\ref{tab:vo-ve-vs-proposed}, our proposed model (utilizing the appended-keyword \textit{speech}) consistently outperforms the benchmark~\cite{lee25e_interspeech} in terms of micro-averaged $\mathrm{\textit{F}}_1$ score (e.g., 0.7625 vs. 0.7286 at optimal thresholds).
This result is particularly notable considering that the benchmark model is explicitly optimized for a fixed, closed-set classification task. In contrast, by leveraging the semantically rich information within LLM embeddings, our framework not only enables open-set prediction but also demonstrates superior discriminative power even within the closed-set domain. This suggests that aligning acoustic features with a structured semantic manifold provides a more robust representation of speaker characteristics than traditional categorical index mapping.

\subsection{Zero-Shot Generalization via Synonym Prediction}
To evaluate the generalizability of our proposed framework, it is essential to validate its performance on attributes not encountered during training. However, as LibriTTS-P is currently the only available dataset with speaker-wise attribute labels, we conduct this validation through a synonym prediction task. Specifically, we transformed the original attribute labels into their respective synonyms using Gemini 3.1 Pro~\cite{deepmind2026gemini31}, accessed in February 2026.

To rigorously assess true semantic understanding rather than simple lexical overlap, we intentionally excluded trivial morphological variants sharing the same lexical root (e.g., simple suffixations like \textit{-ly}). The resulting set of diverse synonyms is detailed in Table~\ref{tab:attribute_synonyms}. While this setup provides a specific view of generalizability, it represents a challenging zero-shot scenario that conventional closed-set classification models are inherently unable to perform. By mapping acoustic features to a continuous semantic space rather than categorical indices, our model demonstrates the capability to recognize attributes through their semantic proximity. 

The results are summarized in Table~\ref{tab:transposed-f1}, where we report performance using various keywords appended to the attributes: {\textit{speech, voice, face, man, apple}}. Notably, the $F_1$ scores across all keyword variants are nearly identical to those observed in the closed-set evaluation (Table~\ref{tab:vo-ve-vs-proposed}). This indicates that our proposed method generalizes effectively to zero-shot attributes, maintaining high accuracy through semantic proximity in the continuous embedding space.
A critical finding is that without the keyword-appending strategy (\textit{No keyword}), performance collapses at high threshold values ($\tau = 0.8$), suggesting that the embedding space settles into an unstable manifold. 

Interestingly, the effectiveness of keyword appending appears to be somewhat independent of the keyword's semantic domain. While \textit{speech} and \textit{voice} are domain-relevant, and \textit{face} or \textit{man} are frequently co-located with attributes in natural language (e.g., ``bright face''), the \textit{apple} keyword—which lacks any semantic association with speaker attributes—also yielded robust performance, even achieving the highest $F_1$ score at $\tau = 0.8$. This implies that the benefit of the keyword-appending strategy may not solely stem from domain-specific semantic guidance, but rather from a structural regularization effect on the LLM embedding space. We further investigate this phenomenon through a geometric analysis in the following section.

\begin{table}[t]
\centering
\caption{Geometric analysis of the attribute embedding space and corresponding performance gains ($\Delta \mathrm{\textit{F}}_1$) across different keyword settings.}
\label{tab:spatial_analysis_combined}
\resizebox{\columnwidth}{!}{%
\begin{tabular}{lccccc}
\toprule
% Variance 부분도 다른 컬럼들과 동일한 tabular 구조로 통일했습니다.
\multirow{2}{*}[-0.5ex]{\textbf{Keyword}} & 
\multirow{2}{*}[-0.5ex]{\begin{tabular}[c]{@{}c@{}}\textbf{Center}\\[-0.3ex]\textbf{Sim}\end{tabular}} & 
\multirow{2}{*}[-0.5ex]{\begin{tabular}[c]{@{}c@{}}\textbf{Total}\\[-0.3ex]\textbf{Variance}\end{tabular}} & 
\multirow{2}{*}[-0.5ex]{\begin{tabular}[c]{@{}c@{}}\textbf{PCA}\\[-0.3ex]\textbf{Log-det}\end{tabular}} & 
\multicolumn{2}{c}{\textbf{$\Delta \mathrm{\textit{F}}_1$ (w/ vs. w/o $\mathcal{L}_{\text{negk}}$)}} \\
\cmidrule(lr){5-6}
& & & & \textbf{Closed-set} & \textbf{Synonym} \\
\midrule
No keyword      & $0.7385$ & $0.4546$ & $-64.221$ & - & - \\
\textit{speech} & $0.8557$ & $0.2678$ & $-87.488$ & $+0.0503$ & $+0.0530$ \\
\textit{voice}  & $0.8531$ & $0.2722$ & $-86.516$ & $+0.0513$ & $+0.0598$ \\
\textit{face}   & $0.8522$ & $0.2737$ & $-86.778$ & $+0.0245$ & $+0.0315$ \\
\textit{man}    & $0.8392$ & $0.2958$ & $-83.950$ & $+0.0079$ & $+0.0005$ \\
\textit{apple}  & $0.8440$ & $0.2876$ & $-84.411$ & $+0.0002$ & $-0.0187$ \\
\bottomrule
\end{tabular}%
}
\vspace{-4mm}
\end{table}

\subsection{Geometric Analysis of the LLM Embedding Space}
To investigate how keyword-appending structurally influences the embedding space, we analyze the manifold geometry using several quantitative metrics. Table~\ref{tab:spatial_analysis_combined} summarizes the results across the following measures:
\begin{itemize}
\item \textbf{Center Sim}: The average cosine similarity between each keyword-appended attribute embedding and its respective centroid. A higher value indicates that the embeddings are more tightly clustered around their semantic centers.
\item \textbf{Total Variance}: The trace of the covariance matrix of the embeddings. A lower value signifies a more compact and constrained embedding distribution.
\item \textbf{PCA Log-determinant (Log-det)}: A metric used to quantify the volumetric cohesion of the embedding manifold. It is defined as the sum of the logarithms of the singular values obtained from the Singular Value Decomposition (SVD) of the zero-centered embedding matrix. A lower value indicates a more compact and narrower embedding space.
\end{itemize}
As shown in Table~\ref{tab:spatial_analysis_combined}, any keyword-appending strategy consistently leads to a geometric contraction of the embedding space. This structural narrowing is likely the primary factor contributing to the stable performance observed at high thresholds ($\tau = 0.8$) in Tables~\ref{tab:vo-ve-vs-proposed} and \ref{tab:transposed-f1}.

Furthermore, we analyze the performance gain ($\Delta F_1$) attributed to the top-$k$ negative loss ($\mathcal{L}_{\text{negk}}$). While positive gains are observed for \textit{speech, voice,} and \textit{face}, the effect is negligible or even slightly negative for \textit{man} and \textit{apple} (e.g., $\Delta F_1\!=\!-0.0187$ for \textit{apple} in the synonym task). We ascribe this trend to the relative density of the resulting manifolds. The spaces formed by \textit{man} and \textit{apple} are relatively wider (higher Log-det) than those of other keywords, providing sufficient inherent separability for acoustic-to-textual mapping. In contrast, \textit{speech, voice,} and \textit{face} induce a significantly narrower manifold. In such crowded semantic spaces, cross-modal alignment becomes more susceptible to interference from proximal negative attributes. Consequently, $\mathcal{L}_{\text{negk}}$ becomes more effective by explicitly penalizing these crowded regions and enforcing discriminative margins.

These findings suggest that the efficacy of speaker attribute prediction is closely related to the compactness of the target manifold. By strategically utilizing keyword-appending and top-$k$ negative penalization, we can effectively structure a manifold that balances semantic cohesion with discriminative clarity. Finally, the hyperparameters for $\mathcal{L}_{\text{negk}}$ were determined through an ablation study, as detailed in Table~\ref{tab:negk_hyperparams}.

\begin{table}[t]
\centering
\caption{Ablation study on the hyperparameters of $\mathcal{L}_{\text{negk}}$: loss weight ($\lambda$), margin ($m$), and the number of negatives ($k$).}
\label{tab:negk_hyperparams}
\resizebox{\columnwidth}{!}{%
\begin{tabular}{ccccccc}
\toprule
\multirow{2}{*}{\textbf{Varying}} & \multicolumn{3}{c}{\textbf{Hyperparameters}} & \multicolumn{2}{c}{\textbf{Micro $\mathrm{\textit{F}}_1$}} \\
\cmidrule(lr){2-4} \cmidrule(lr){5-6}
& Weight ($\lambda$) & Margin ($m$) & Negative ($k$) & \textbf{Closed-set} & \textbf{Synonym} \\
\midrule
\textbf{Default} & $0.5$ & $0.2$ & $1$ & $0.7380$ & $0.7450$ \\
\midrule
\multirow{2}{*}{$\lambda$} & $0.1$ & $0.2$ & $1$ & $0.7340$ & $0.7378$ \\
 & $1.0$ & $0.2$ & $1$ & $0.7201$ & $0.7200$ \\
\midrule
\multirow{2}{*}{$m$} & $0.5$ & $0.1$ & $1$ & $0.6479$ & $0.6323$ \\
 & $0.5$ & $0.3$ & $1$ & $0.6745$ & $0.6590$ \\
\midrule
\multirow{2}{*}{$k$} & $0.5$ & $0.2$ & $5$ & $0.6587$ & $0.6687$ \\
 & $0.5$ & $0.2$ & $10$ & $0.7273$ & $0.7255$ \\
\bottomrule
\end{tabular}%
}
\vspace{-4mm}
\end{table}

\section{Limitations and Conclusion}
In this work, we proposed a novel framework for open-set speaker attribute prediction using LLM-based semantic embeddings. By introducing a keyword-appending strategy and employing top-$k$ negative penalization, we effectively structured a discriminative semantic manifold that bridges the cross-modal gap between audio and text. 
% Experimental results on LibriTTS-P demonstrated that our approach not only outperforms closed-set benchmarks but also generalizes to unseen synonyms with high precision. 
Experimental results on \mbox{LibriTTS-P} demonstrated that our approach not only outperforms closed-set benchmarks but also exhibits robust generalization to unseen synonyms.
Despite these contributions, this study has limitations. While we prioritized LibriTTS-P due to its absolute attribute labels—unlike the relative comparisons in VCTK-RVA~\cite{sheng2025voice}—the reliance on a single corpus may limit broader stylistic validation. Furthermore, while we utilized GPT-OSS-20B as a representative LLM, investigating how different LLM architectures and embedding spaces influence manifold geometry remains a task for future work. We expect our manifold-structuring approach to serve as a foundation for more interpretable and extensible speaker characterization systems.

\section{Acknowledgements}
This work was supported by Basic Science Research Program through the National Research Foundation of Korea (NRF) funded by the Ministry of Education [No. RS-2025-25398143, 50\%], National Research Foundation of Korea (NRF) grant [No. RS-2025-24683892, 45\%] and Institute of Information \& communications Technology Planning \& Evaluation (IITP) grant [No. RS-2021-II211343, 5\%] funded by the Korea government (MSIT). The GPUs were supported by the Advanced GPU Utilization Support Program funded by the Ministry of Science and ICT and supervised by NIPA [No. 02-26-01-0285].

\section{Use of Generative AI Disclosure}
The authors used generative AI tools only for paraphrasing and wording refinement to improve the readability and completeness of the manuscript. The authors have reviewed the manuscript and take full responsibility for its content.

\bibliographystyle{IEEEtran}
\bibliography{mybib}

\end{document}